# Modern Design Methodologies and the Development of Mechatronic Products

Prof. Rogerio Atem de Carvalho, D.Sc.

Innovation Hub, Instituto Federal Fluminense, Brazil

This article presents a quick view on the development of mechatronic products and how the techniques of Design Thinking, Concurrent Engineering and Agilism can be integrated to address this development. Design Thinking is employed in the early stages in order to better explore creativity, whereas Concurrent Engineering and Agilism are applied during the development of the product, in order to deal with emerging requirements, typical of the development of complex products.

## 1. Introduction

The current economic environment is characterized by increasing customer requirements on product performance, quality, and price, leading to a highly dynamic product design environment with shortened development and product lifecycles. Product innovation then participates in a significant way in obtaining and maintaining the competitive capacity of companies. In this scenario, dealing with the design of innovative products becomes a challenge, given that this activity usually demands the solution of technological challenges while dealing with emerging product requirements.

This scenario also favors the growth of the application of the concepts of Mechatronics, which unites the principles of mechanics, electronics, and computing to generate simpler, more economical, and reliable products and manufacturing systems, making the division between these disciplines blurred, demanding for more integrated approaches to engineering in general.

In this scenario, this article intends to briefly discuss the background of the development of mechatronic products and to address the main aspects of the dominant modern design methodologies, presenting in which points they converge and in which they diverge, in order to give an initial direction that allows the development of mechatronic product design processes based on them. In order to achieve this, it is structured as follows, after this Introduction:

-Topic 2 is a brief review on developing mechatronic products;

-Topic 3 is a presentation of the main features of the selected design methodologies;

- Topic 4 introduces the convergent and divergent points of the methodologies;

- Topic 5 presents the the conclusions.

## 2. Design of Mechatronic Products

Although the concept of Mechatronics refers back to the early 1970s, it should be considered from the year 2000 onwards that there was a great leap forward in Information Technology, and hence, from then on Mechatronics also becomes even more present in society in general. Considering this recent period then, it can be stated that the VDI[1] 2206 standard, "Design Methodology for Mechatronic Systems", of the German Association of Engineers, is the one that most influenced the most followed in the area of the definition of mechatronics product design processes. The VDI 2206 started to be defined in the year 2000 and was created based on experiences of the industrial practice and the results of empirical design research, which show that there is no fixed form of an optimal process that an engineer can follow. Thus, according to Gausemeier and Moehringer (2002), the VDI 2206 is a flexible procedure model, based on three main elements:

- A general cycle of problem-solving on the micro-level;

- The V-Shaped model on the macro-level, and

- Predefined process modules for repeating operation steps during the design.

---

[1] Verein Deutscher Ingenieure, the Association for Germans Engineers.

The first and third elements above come from Systems Engineering, while V-Shape (or V-Model) comes from Software Engineering, and has been adapted to mechatronic design needs. That is, the VDI 2206 is a meta-process that integrates established techniques in favor of a coherent development between the three disciplines considered: Mechanics, Electronics and Computing. After the diffusion of the VDI 2206, in the last decade and a half a series of other processes have been proposed, aiming to adapt it to specific circumstances or to improve it. At this point, articles of relevance that explicitly address changes in the macro-process based on the V-Shape (or V-Model) in recent years are commented below.

Chami and Bruel (2015) state that Mechatronics has been considered as one of the main innovation leaders in the industry, and that in each of its design and development process phases, a wide range of languages, methods, and tools are used. Chami and Bruel (2015) also state that one of the main problems in this area is the integration, for instance when changes in a given requirement are needed, the engineers "have to start analyzing from this requirement, moving forward to the refined functions, concepts, and proposed solutions while considering all discipline specific criteria. Consequently, the complexity arises due to the fact that this entities level is directly dependent on the other factors such as the tools, methods and human factors levels." Highlighting that system engineers still follow a document-based approach to hold the disciplines' interdependencies, which leads to weak synchronization and result in inefficiencies that often appear during the integration or testing. In order to solve this and other problems, several methodologies have been developed, with a lot in common, nevertheless, the literature agrees that there is no "one accepted methodology" and from another practical side, this problem seems to be hardly solved as companies are individually developing their own methodologies. Chami and Bruel (2015) explore the application of Model-Based Systems Engineering and Artificial Intelligence for supporting the mechatronics design, introducing SysDICE, a method for making use of SysML as a common language between system and discipline engineers.

Salem and Mahfouz (2016) deal with Mechatronic systems design, also considering educational aspects. According to this view, Mechatronics system design is a modern interdisciplinary design procedure, comprised of concurrent selection, evaluation, synergistic integration and optimization of the whole system and all its sub-systems and components as a whole, with all the design disciplines working in parallel and collaboratively throughout the design and development process to produce an overall optimal design. This approach offers fewer constraints and shortened development, also allows the design engineers to provide feedback to each other about how their part of the design is effect by others. here it is important to notice that the ideal process of concurrent (simultaneous) approach is characterized by parallel work of a potentially distributed community of designers that know about the parallel work of their colleagues and collaborate as necessary; sharing of knowledge in a common database builds a basis of cooperative design, since a shared database is the place where all design results are integrated. In this context, Salem and Mahfouz (2016) propose a guideline comprised of a series of steps:

- Pre-study – Problem Statement;

- Conceptual Design;

- Modeling, Simulation, Analysis, and Evaluation;

- Prototyping, Testing, Evaluation, and Optimization;

- Manufacturing and Commercialization;

- Support, Service, and Market Feedback Analysis.

These steps aim at guaranteeing not only the coverage of all the activities necessary to the design of new products as well as to build knowledge around the engineering process.

Zheng et al. (2017) present a concurrent, hierarchical design methodology, divided into a macro level and micro level. The macro-level process is used to describe the generic procedure for the design of mechatronic systems, and it can be specified according to the individual design phase which is called the micro-level process. The micro-level process supports every specific design phase where individual designers can structure design sub-tasks and proceed and react in unforeseen situations. The proposed design methodology adopts an extended V-model as the macro-level process and the hierarchical design model as the micro-level process. In particular, Zheng et al. (2017) extend the V-Shape by introducing a decision point after the Integration Tests. A multidisciplinary interface model is proposed to ensure the consistency and traceability between the two levels. In the extended V-model, the right branch of the V-model represents the system integration sub-process, which is divided into two phases: Compatibility test phase and Verification and validation phase. The objective of the Compatibility test phase is to guarantee the subsystems integrate correctly and to ensure the multidisciplinary integration among different design teams. If the sub-systems are incompatible with each other, an iterative process is carried on. This compatibility test in the early phase of system integration greatly reduces the number of iterations in the later phase so that the overall development cost and the time-to-market decrease accordingly. A Verification and Validation phase is used to test the performance of the integrated system and check whether the system realizes the proposed function and satisfies all the proposed requirements.

Eigner, Gilz and Zavirov (2018) present a model-based, data integration process for the interdisciplinary development of products, focusing on the left wing of the V-model and extending it. Eigner, Gilz and Zavirov (2018) state that a complete definition of requirements, functions and logical system elements, (the three proposed views), is difficult to achieve directly from the start, therefore, the development process should follow the V-shape going in incremental loops that refine all aspects further. Their process is based on functional and data models. The functional product description model is divided into Hierarchies, Cross-references between model elements of different types, and Typed Internal Connections among model elements of the same type. The data model for product data management searches to integrate the functional model to the Product Lifecycle Management activities, assuming the Product Design Management system already supports the management of requirement and BOM structures. The data model integrates the functional product description between the requirements and the BOM, modeling functions as hierarchical structures, which in turn can be modeled as block definitions in SysML editors. According to Eigner, Gilz and Zavirov (2018), the data model facilitates the tracking of complex relationships within the products and their interaction with each other, especially in the early phase of the development process, thus better integrating the design of the products to the Enterprise Resource Planning (ERP) functions, such as customer and supplier management (CRM and SRM).

As can be seen, the studies presented here propose derivations of the V-shape through different approaches, while drawing similar conclusions: there is no optimal process adaptable for all situations, the development activities of the different disciplines need to occur concurrently, and the aspects of the product should be refined through iterations. In this sense, and without intending to propose another derivation of the V-Shape, the process presented here focus on the development of innovative mechatronic products and is situated at a higher level of abstraction, when compared to the others briefly described previously. This said different models of micro-management can be adapted to the proposed process.

**3. Modern Design Methodologies**

Best practices for the design of products can be found on the way project-based organizations operate. According to Hölzle and Holger (2019), in these organizations, the majority of products or services are produced through projects for either internal or external customers. Following this direction, it is necessary to investigate the methodologies that help project-based organizations to execute their projects efficiently within the economic scenario enunciated at the beginning of this article. There are currently three methodologies that stand out in terms of positive impacts in this area: Concurrent Engineering, Agilism, and Design Thinking.

Concurrent Engineering (CE) is a design methodology that has had the opportunity to mature in recent years to become a well-defined systems approach towards optimizing engineering design cycles of complex systems (Ma et al., 2008). Because of this, CE has been implemented in a number of organizations, most notably in the aerospace industry. Among the CE premises is that the preceding design activities should all be occurring at the same time, i.e., concurrently. The idea is that the concurrent nature of these processes significantly increases productivity and product quality. This way, errors, and redesigns can be discovered early in the design process when the project is still flexible. By locating and fixing these issues early, the design team can avoid what often turns into costly errors, as the project moves to more complicated models and eventually into the actual manufacturing of hardware of a complex system (Kusiak, 1993).

Talking about modern complex systems means talking about software, whose role within systems, in general, has fundamentally changed over the past 50 years. According to Sheard (2014), the software's role has changed both in mission-critical systems, such as fighter aircraft and medical equipment, to commercial products, such as mobile telephones and consumer electronics. The software has become not only the brain of most systems but also the backbone of their functionality. In the mechatronics arena, for instance, software applications, such as those for controlling devices and instruments, have been growing rapidly since the 1980s. The demand for producing software with small memory and processor footprints and high reliability has grown even more, given the emergence and exponential growth of Internet of Things (IoT) applications, products, and systems.

However, what does developing good embedded software mean? In the mechatronics area, good software means embedded software with small memory and processor footprints and high reliability. In this sense, it has been known for more than a decade that the way to producing good embedded systems inevitably passes through the use of integration and testing "at the very fabric of software development", as advocates Ganssle (2008) and using automated tests into this fabric (GRENNING, 2011, P. 2). Integration and automated testing leave at the core of Agile software development, which is a group of methods in which solutions evolve through collaboration between self-organizing, cross-functional teams. Agilism promotes adaptive planning, evolutionary development, early delivery, continuous improvement, and encourages rapid and flexible response to change (Agile Alliance, 2013). Given the similar philosophies followed by Concurrent Engineering and Agile Methods, it can be said that both can be combined in the development of mechatronic products, or the second can be integrated into the first in order to explore synergies. Hemphälä (2014) for instance identified that Agile Methods represent a new addition to the embedded systems development toolbox, together with Concurrent Engineering. In fact, the principles of both techniques can be traced back to the Lean Thinking school, indicating that integration is reasonably natural.

While Concurrent Engineering and Agile Methods serve well the incremental design and detailing requirements of the product, what about the development of innovative products, where most of the times the requirements themselves are not well defined, making it difficult for these design and detailing activities start and develop at an acceptable risk level? Gartzen et al. (2016) state that innovative products are general is characterized by high degrees of both market and technological uncertainty, and the necessary reduction of the time to market and the development costs can be achieved by parallelization and forward displacement of activities, as with Concurrent Engineering and front-loading. In the case of more radical innovations, however, Gartzen et al. (2016) state that a learning-oriented approach is recommended, with a high degree of flexibility and agility to give development teams more freedom of creativity, by developing incrementally physical products through prototypes, corresponding to the development of functional code for software, which allows the team to learn and incorporate obtained information into the next development cycles. Gartzen et al. (2016) then propose a methodical approach to prioritize items from the design backlog, aiming at following the most efficient development path through the prototypes. The potential customers and other stakeholders can be integrated into the early stages of the development process, in order to provide important information about their needs, including the latent, and requirements. Additionally, prototypes represent milestones of the process, acting as a channel of communication between stakeholders.

Prototyping is also a core feature of Design Thinking, which is a user-centric set of design methods and principles that first appeared in the 1980s and was developed and made popular by David Kelley and Tim Brown over the late 1990s (Kelley–Littman 2001), and it is prescribed for the design of innovative products and services (Brown, 2008). As part of a design process, Design Thinking is well suited to approach will-defined requirements, encompassing a series of techniques among which stands out:

- Problem framing, which means that rather than accept the problem as given, designers explore the given problem and its context in order to reach a particular framing of the problem that points to a path to a solution (Dorst, 2011);

- Solution-focusing thinking, instead of problem-focused thinking;

- Co-evolution of problem–solution, which means exploring the problem and its solution in parallel, leading to a deeper understanding of the core problem, which in turn may help to find additional solution components (Wiltschnig; Christensen; Ball, 2013).

Using these techniques and more, Design Thinking is ideal for brainstorming in the early stages of developing innovative products, however, it is also criticized for oversimplifying the design process, undermining the importance of the technical skills (Kolko, 2018).

Hence, in the design of innovative products, while Design Thinking is ideal for exploring ideas and requirements in the early stages of a product development project, Concurrent Engineering and Agilism are more adequate to coordinate the developers' technical skills in the more advanced phases of the design. What is desired is to integrate these methodologies into a high-level process, which can be derived for specific organizations and problems.

**4 Integrating the Methodologies**

In order to integrate the three methodologies, it is necessary to evaluate at which points they converge and at which they diverge.

*4.1 Convergent Practices*

Teamwork: The very basis of the three techniques, the design team is divided into cells, each of Computing, Electronics, and Mechanics. For bigger projects, each discipline can have more than one cell. Cells can be arranged in different ways, including as a single team, especially during brainstorming tasks, or operating individually as discipline-specific tasks emerge. Additionally, during the integration iterations, the cells tend to be merged into bigger cells, or, again, as a single team.

Knowledge Sharing: Direct consequence of the teamwork, it occurs in the intra-discipline form, when the work is performed by the individual cells, and in the inter-discipline form when the cells join.

Decision Sharing: This practice is implemented by two others, namely: "one person, one vote", which means that design decisions are, in most cases, taken in groups, where each individual has equal rights; and "the most able leads", in the sense that cell leader can change depending on the specific task being performed, and the person with the biggest technical expertise will then lead the task execution. The task leader deliberates with the team and decides through consensual decisions, in the cases of greater complexity. The idea is to avoid monocratic decisions so that everyone is responsible for the paths chosen for the project.

Flexibility to Change: Developing innovative products means dealing with emerging requirements, so you need to be able to make changes to the product quickly and at acceptable costs. The heavy use of automated testing for software and hardware, and prototyping for hardware and mechanical components is intended to provide this capability.

Task Independence: Cell activities should be as independent as possible from each other. Basically, it seeks to clearly define the interfaces between the disciplines, and consequently between the software, hardware, and mechanical components, and to establish (micro) contracts that ensure that the components will work in an integrated manner. In this way, it is possible for each cell to perform its work in parallel to another, reducing the total time, while ensuring systemic coherence.

Early Error Detection: Agilism and Concurrent Engineering use constant implementation, integration, and testing in order to detect errors as soon as possible while Design Thinking trusts in prototyping in the way of constantly checking the functioning of the product as it evolves. dIP uses prototyping together with automated tests to ensure early error detection.

Design for Sustainability: This practice is explicitly referred in Concurrent Engineering and Design Thinking, in Agilism however, the reference is implicit, since it aims at developing the product meets the requirements and only the requirements, in the most efficient way. dIP seeks to stimulate the stakeholders to think about the issues that involve the decommissioning of the product from its conception.

*4.2 Divergent*

The time regarding technical Decision Making is, basically, the single divergence among the three techniques. Agilism uses Late Decision Making – from its origin in the software industry and the ability to more easily change the product; Concurrent Engineering uses Early Decision Making, from its origin in the aerospace industry and its demand for avoiding late and costly changes; Design Thinking relies on prototyping and while uses early decision making for the high-level requirements, in theory, adopts late decision making in the technical arena (while evolving the prototypes towards the final product). In order to mitigate this divergence among the techniques, dIP uses a cost-based logic to determine when to freeze certain characteristics of the product: the higher the estimated cost of change, the sooner the decision has to be taken, freezing the technical characteristic and deriving the others in function of it. Although the use of prototyping pushes the decision-time window forward in general, software-related decisions tend to be delayed, intermediate timing is used for hardware, and early timing for mechanical components.

## 5. Conclusions

The majority of the current studies in mechatronic product development propose derivations of the V-shape and draw similar conclusions, namely: there is no such a process adaptable to all situations, the development activities of the different disciplines need to occur concurrently, and the aspects of the product should be refined through iterations. This paper contributed by, following these findings, investigating the most influential techniques of the present time, and introducing the first steps aiming to compose a specific process for the development of innovative mechatronic products.

The main conclusions is that, while most of the techniques employed by the three methodologies are convergent - clearly showing that they can be integrated, the main point of divergence is the timing for making design decisions. The three techniques diverge a lot, with Agilism deciding late, Concurrent Engineering deciding early and Design Thinking counting on experimentation to decide. Thus, any process that aims to integrate these methodologies will have to clearly establish which phases will be employed or which decision-making process should be followed. Starting with the identified points of convergence and divergence, it is then possible to define an integrated process for the design of mechatronic products.

**Funding**

This work was sponsored by the National Council for Research (CNPq), under the Technological Development grant 308726/2016-2; the Brazilian Association for Industrial Research & Innovation (EMBRAPII), under the development project PIFF-1803.0009; and the Secretary for Technological and Professional Education (Setec) of the Ministry of Education of Brazil (MEC), under the public call 01/2017, for developing innovative solutions for the Ministry.

**References**

Agile Alliance. What is Agile Software Development? 2013. Retrieved from http://www.agilealliance.org/the-alliance/what-is-agile/ in 05 August 2015.

Brown, Tim. "Design thinking." Harvard business review 86, no. 6 (2008): 84.

Chami, Mohammad, and Bruel, Jean-Michel. "Towards an integrated conceptual design evaluation of mechatronic systems: The SysDICE approach." *Procedia Computer Science* 51 (2015): 650-659.

Dorst, K. (2011) "The Core of Design Thinking and its Application", *Design Studies*, **32**, 521-532.

Eigner, M, Gilz, T., Zavirov, R. Interdisciplinary Product Development - Model Based Systems Engineering, https://www.plmportal.org/en/research-detail/interdisciplinary-product-development-model-based-systems-engineering.html Acessed in 17/05/2018.

Gartzen, T; Brambring, F; Basse, F. Target-oriented prototyping in highly iterative product development. Procedia CIRP 51 (2016): 19-23.

Gausemeier, J.; Moehringer, S. "VDI 2206-A new guideline for the design of mechatronic systems." IFAC Proceedings Volumes 35, no. 2 (2002): 785-790.

Grenning, J. W. Test Driven Development for Embedded C, Pragmatic Programmers, 2011.

Hemphälä, J. An Investigative Eye on the Agile Manifest, 2014. Retrieved from https://www.kth.se/en/itm/inst/mmk/om/nyheter/an-investigative-eye-on-the-agile-manifest-1.470936 in 05 December 2018.

Hölzle, K., and Holger R. "The Dilemmas of Design Thinking in Innovation Projects." Project Management Journal 50, no. 4 (August 2019): 418–30.

Kelley, K. and Littman, J. (2001). The Art of Innovation: Lessons in Creativity from IDEO, America's Leading Design Firm. New York: Doubleday.

Kolko, J. "The divisiveness of design thinking." ACM Interactions, May-June, 2018: http://interactions.acm.org/archive/view/may-june-2018/the-divisiveness-of-design-thinking . Retrieved in 05/02/2019.

Kusiak, Andrew; Concurrent Engineering: Automation, Tools and Techniques. John Wiley & Sons, 1993.

Ma, Y., Chen, G., Thimm, G. "Paradigm Shift: Unified and Associative Feature-based Concurrent Engineering and Collaborative Engineering", Journal of Intelligent Manufacturing, DOI 10.1007/s10845-008-0128-y

Salem, F. A., Mahfouz, A. A. Mechatronics Subsystems' Classification, Role, Selection Criteria and Synergistic Integration in Overall System Design, American Journal of Educational Science, Vol. 2, No. 3, 2016, pp. 16-28.

Sheard, S. Needed: Improved Collaboration Between Software and Systems Engineering. http://blog.sei.cmu.edu/post.cfm/improved-collaboration-software-systems-engineering-139?utm_content=bufferc7d75&utm_medium=social&utm_source=linkedin.com&utm_campaign=buffer Software Engineering Institute, 2014.

Sheard, S. The Changing Relationship of Systems and Software in Satellites: A Case Study. https://blog.sei.cmu.edu/post.cfm/changing-relationship-systems-software-satellites-case-study-209. Software Engineering Institute, 2014.

Wiltschnig, Stefan; Christensen, Bo; Ball, Linden (2013). "Collaborative problem–solution co-evolution in creative design". Design Studies. 34 (5): 515–542. doi:[10.1016/j.destud.2013.01.002](https://doi.org/10.1016/j.destud.2013.01.002).

Zheng, C., Hehenberger, P., Le Duigou, J., Eynard, B. Multidisciplinary design methodology for mechatronic systems based on interface model. Res Eng Design (2017) 28:333–356.